\documentclass[aps, prd, groupedaddress, preprint, tightenlines,
  eqsecnum, nofootinbib, showpacs]{revtex4}

\usepackage{axodraw, float, slashed, graphicx, amssymb, amsmath}
\usepackage[usenames, dvipsnames]{xcolor}

\usepackage[margin=2.2cm, a4paper, includefoot]{geometry}

\def\caja{\mathsurround=0pt}
\def\eqalign#1{\,\vcenter{\openup1\jot \caja
        \ialign{\strut \hfil$\displaystyle{##}$&$
         \displaystyle{{}##}$\hfil\crcr#1\crcr}}\,}
\numberwithin{equation}{section}

\renewcommand{\theequation}{\thesection.\arabic{equation}}
\def\equn{\refstepcounter{equation}\eqno({\rm \theequation})}

\def\npb#1#2#3{{\rm Nucl. Phys. B}{\bf \ #1}, #3 (#2)}
\def\spa#1.#2{\left\langle#1\,#2\right\rangle}
\def\spb#1.#2{\left[#1\,#2\right]}
\def\eps{\epsilon}
\def\la{\langle}
\def\ra{\rangle}

\def\Atree{A^{\tree}}
\def\Aloop{A^{\oneloop}}
\def\BR#1#2{[#1|{K_{abc}}|#2\ra}
\def\BRTTT#1#2{\la#1^+|\slashed{K}_{abc}|#2^+\ra}
\def\NeqEight{\Neq8}
\def\NeqSix{\Neq6}
\def\NeqFour{\Neq4}
\def\NeqOne{\Neq1}
\def\NeqZeroF{\Neq0}
\def\NeqZero{[0]}
\def\deff{d_{\text{eff}}}
\newcommand{\oneloop}{\text{1-loop}}

\newcommand{\tree}{\text{tree}}
\newcommand{\Neq}[1]{\mathcal{N} = #1}

\DeclareMathOperator{\Perm}{\mathcal{P}}

\DeclareMathOperator{\tr}{\mathrm{tr}}
\DeclareMathOperator{\SP}{\mathrm{Split}}
\DeclareMathOperator{\Soft}{\mathrm{Soft}}
\newcommand\trunc{\text{trunc}}
\newcommand\figref[1]{fig.~\ref{#1}}
\def\e{\epsilon}

\begin{document}

\title{Obtaining One-loop Gravity Amplitudes Using Spurious Singularities}

\author{David~C.~Dunbar, James~H.~Ettle and Warren~B.~Perkins}

\affiliation{Department of Physics, \\
College of Science, \\
Swansea University, \\
Swansea, SA2 8PP, UK}

\begin{abstract}
 
  The decomposition of a one-loop scattering amplitude into elementary
  functions with rational coefficients introduces spurious
  singularities which afflict individual coefficients but cancel in
  the complete amplitude.  These cancellations
  create a web of interactions between the various terms.  We explore
  the extent to which entire one-loop amplitudes can be determined from these
  relationships starting with a relatively small input of initial
  information, typically the coefficients of the scalar integral functions as these
are readily determined. 
In the context of one-loop gravity
  amplitudes, of which relatively little is known, we find that some
  amplitudes with a small number of legs can be completely determined
  from their box coefficients. For increasing numbers of legs,
  ambiguities appear which can  be determined from the
  physical singularity structure of the amplitude. We illustrate this
  with the four-point and $\Neq1,4$ five-point (super)gravity
  one-loop amplitudes.

\end{abstract}

\maketitle

\section{Introduction}

In general, the one-loop amplitudes of a quantum field theory can be
expressed as a sum over Feynman diagrams
$$
\Aloop_n=\sum_{a\in \cal L}\, D_a \equn\label{FeynmanBasis}
$$ 
where the summation is over all diagrams constructible from the
vertices and propagators of the theory. In a gauge theory the number
of diagrams grows exponentially with the number of external legs.  The
one-loop diagrams will involve the integral over a loop momentum
$$
D_a = \int d^D\ell { P(\ell) \over \prod (\ell-K_i)^2 } \equn
\label{BasicLoopEq}
$$
where the $(\ell-K_i)^{-2}$ are the propagators attached to the loop
and $P(\ell)$ is a polynomial of Lorentz invariants constructed by
contracting the loop momentum with the momenta and polarisations of
the external states. If $r$ is the number of propagators in the loop,
in a Yang--Mills theory $P(\ell)$ is a polynomial of degree $r$ and in
a gravity theory $P(\ell)$ is a polynomial of degree $2r$.  The
integrals are regularised by calculating with $D=4-2\eps$.

Powerful and well-established integral reduction methods
\cite{IntegralReduction} allow an $r$-point, rank-$p$ one-loop integral to
be expressed as a sum of $r-1$ point integrals, to $O(\eps)$,
$$
I_r[ P^{p}[\ell] ] =\sum d_i I^{(i)}_{r-1}[P^{p-1}[\ell] ], \quad r >
4. \equn
$$ 
The set of diagrams in the composition is those where one propagator
of the parent is collapsed.  For $r=3,4$ the decomposition is
$$I_r[ P^{p}[\ell] ] 
= I_r[1]+\sum d_i I^{(i)}_{r-1}[P^{p-1}[\ell] ] \equn
\label{ReductionEquation}
$$ 
and for $r=2$,
$$
I_2[ P^{p}[\ell] ] = I_2[1]+R \equn
$$
where $R$ is a rational function of the Lorentz invariants.  The end
result is that any amplitude in a massless theory can be expressed as
$$
\Aloop_n=\sum_{i\in \cal C}\, a_i\, I_4^{i} +\sum_{j\in \cal D}\,
b_{j}\, I_3^{j} +\sum_{k\in \cal E}\, c_{k} \, I_2^{k} +R_n +O(\eps),
\equn\label{DecompBasis}
$$ 
where the $I_r$ are $r$-point scalar integral functions and the $a_i$
etc. are rational coefficients. $R_n$ is a purely rational term.

Dividing the amplitude into integral functions with rational
coefficients has been very fruitful: a range of specialised techniques
have been devised to determine the rational coefficients,  many based
on unitarity techniques rather than Feynman
diagrams~\cite{Cutkosky:1960sp,Bern:1994zx,Bern:1994cg,Bern:1997sc,BrittoUnitarity,Bern:2004llygdd,Bidder:2005ri,BBCFsusyone,Darren,BjerrumBohr:2007vu,Mastrolia:2006ki,
NigelGlover:2008ur,UsUnitarity,Bern:2005cq,Berger:2006ci,blackhat,Badger:2008cm}. 
 Progress has been made both via the two-particle cuts and using 
generalisations of unitarity~\cite{Bern:1997sc} where, for example, 
triple~\cite{Bidder:2005ri,Darren,BjerrumBohr:2007vu,Mastrolia:2006ki} and quadruple cuts~\cite{BrittoUnitarity} 
are utilised to identify the triangle and box coefficients respectively.

However there is a cost in the division: the rational
coefficients do not have the symmetries or singularity structure of
the full amplitude. In particular they may acquire ``spurious
singularities''.  In general the Passarino-Veltman reduction
coefficients $d_i$ of \eqref{ReductionEquation} contain a factor of
$\Delta^{-1}$ where $\Delta$ is the Gram determinant of $I_r$.  The
vanishing of $\Delta$ does not necessarily correspond to any physical
singularity of the amplitude.  Such singularities arising from the
reduction must cancel between the various contributions 
to the complete amplitude.  In this article we explore the web of
cancellations which link the ``cut-constructible''  parts of the
amplitude to the rational terms.  This has been explored in
Yang-Mills theories as part of the bootstrapping
process~\cite{Bern:2005cq,Berger:2006ci,blackhat} where the spurious
singularities are combined within integral functions.  Here, the
spurious singularities arising in the ``cut-constructible'' parts of
the amplitude and cancelled by rational terms constructed from the full 
integral coefficients and modifications of the integral functions. 
In gravity we find that the spurious
singularities occur with higher powers which consequentially place
stronger constraints on the rational terms. We also find in the
$\NeqOne$ case,  it is not practical to 
simultaneously combine all
such singularities into integral functions 
leaving the coefficients
unchanged but instead adopt an
approach where we identify the different singularities with a single
coefficient
and cancel these iteratively in a specified order. 

This is useful since  supergravity amplitudes are relatively difficult to
calculate with only a small number of one-loop helicity amplitudes
available for study.  For $\NeqEight$ supergravity the one-loop
structure is relatively well understood:  the expansion is purely in
terms of scalar box integrals as demonstrated in the explicit
calculations of the four-point MHV amplitude~\cite{Green:1982sw}, the
$n$-point MHV amplitude~\cite{MaxCalcsB}  and the six and seven-point
NMHV amplitudes~\cite{MaxCalcsC,BjerrumBohr:2006yw}.  For $\NeqSix$ 
the $n$-point MHV amplitude is also known~\cite{Dunbar:2010fy}.
For ${\cal N} < 6$ very few amplitudes are known. All the four-points amplitudes have
been calculated~\cite{GravityStringBasedB} but only the
five-point MHV $\NeqFour$ is available. For pure gravity, the entirely
rational ``all-plus'' $n$-point amplitude is known~\cite{MaxCalcsB},
the four, five and six-point ``one-minus'' 
and the four-point MHV are known~\cite{Grisaru:1979re,GravityStringBasedB,Dunbar:2010xk}. 

In this article,  we first discuss the physical singularities expected
in amplitudes before illustrating the spurious singularities with
examples in Yang-Mills amplitudes.   We then examine one-loop graviton
scattering amplitudes first with the known examples of scalar four-point and
the five-point $\NeqFour$ amplitudes. Finally we use 
spurious singularity cancellation and physical factorisation to derive the previously unknown $\NeqOne$
five-point amplitude.   Spurious singularities have been used
previously to constrain and determine coefficients in amplitudes in
special cases in  Yang-Mills theories~\cite{Korchemsky:2009jv,BjerrumBohr:2007vu}. 

\section{Physical Singularities of Scattering Amplitudes}

Physical singularities correspond to physical factorisations.  The
factorisation of one-loop massless amplitudes is described in
ref.~\cite{BernChalmers},
\begin{equation}\begin{split}
    \label{LoopFact}
    &A_{n}^{\oneloop} \mathop{\longrightarrow}^{K^2 \rightarrow 0}
    \sum_{\lambda=\pm} \Biggl[ A_{r+1}^{\oneloop}\big(k_i, \ldots,
    k_{i+r-1}, K^\lambda\big) \, {i \over K^2} \,
    A_{n-r+1}^{\tree}\big((-K)^{-\lambda}, k_{i+r}, \ldots,
    k_{i-1}\big) \\ & + A_{r+1}^{\tree}\big(k_i, \ldots, k_{i+r-1},
    K^\lambda\big) {i\over K^2}
    A_{n-r+1}^{\oneloop}\big((-K)^{-\lambda}, k_{i+r}, \ldots,
    k_{i-1}\big) \\
    & + A_{r+1}^{\tree}\big(k_i, \ldots, k_{i+r-1}, K^\lambda\big)
    {i\over K^2} A_{n-r+1}^{\tree}\big((-K)^{-\lambda}, k_{i+r},
    \ldots, k_{i-1}\big) F_n\big(K^2;k_1, \ldots, k_n\big) \Biggr],
  \end{split}\end{equation}
where the one-loop `factorisation function' $F_n$ is
helicity-independent.

When the momentum $K$ consists of just two external momenta,
$K=k_a+k_b$, the limit is subtle because the three-point tree
amplitude vanishes. In a Yang--Mills theory the amplitude has
collinear singularities of the form $\spa{a}.b^{-1} $ and/or
$\spb{a}.b^{-1}$ rather than $s_{ab}^{-1}$.  Note that
$|\spa{a}.b|=|\spb{a}.b|=|s_{ab}|^{1/2}$.  Gravity amplitudes are not
singular in the collinear limit, but take a form that is specified in
terms of amplitudes with one less external leg~\cite{MaxCalcsB}. If
$k_a \longrightarrow z K$ and $k_b \longrightarrow (1-z) K $,
$$
M_n( \cdots , a^{h_a}, b^{h_b} ) \longrightarrow \sum_{h'} \SP_{-h'}
(z,a^{h_a},b^{h_b}) M_{n-1} (\cdots , K^{h'} ) \equn
$$
where the $h$'s denote the various helicities of the
gravitons.\footnote{
The normalisation of the physical amplitude 
${\cal  M}^\tree=(\kappa/2)^{n-2} M^\tree, 
{\cal  M}^\oneloop=(\kappa/2)^{n} M^\oneloop$.}
The
``splitting functions'' are~\cite{MaxCalcsB}
\begin{align}
  \SP_{+}(z,a^+,b^+) &= 0, \\
  \SP_{-}(z,a^+,b^+) &= -{ \spb{a}.{b} \over z(1-z)  \spa{a}.{b}  }, \\
  \SP_{+}(z,a^-,b^+) &= -{ z^3\spb{a}.{b} \over (1-z) \spa{a}.{b} }.
\end{align}
As usual, we are using a spinor helicity formalism with the usual
spinor products $ \spa{j}.{l} \equiv \langle j^- | l^+ \rangle =
\bar{u}_-(k_j) u_+(k_l)$ and $\spb{j}.{l}\equiv \langle j^+ | l^-
\rangle = \bar{u}_+(k_j) u_-(k_l)$, and where $\BR{i}{j}$ denotes
$\BRTTT{i}{j}$ with $K_{abc}^\mu =k_a^\mu+k_b^\mu+k_c^\mu$ etc. Also
$s_{ab}=(k_a+k_b)^2$, $t_{abc}=(k_a+k_b+k_c)^2$, etc.

Gravity amplitudes also have soft-limit singularities~\cite{BerGiKu}
as $k_n \longrightarrow 0$,
$$
M_n( \cdots , n-1, n^h) \longrightarrow \Soft (n^h) M_{n-1} (\cdots ,
n-1 ) \equn
$$
where the ``soft factor'' is given by
$$
\Soft(n^+) =-{ 1\over \spa{1}.n\spa{n}.{n-1} } \sum_{j=2}^{n-2} {
\spa{1}.j  \spa{j}.{n-1} \spb{j}.n\over \spa{j}.n } .  \equn
$$

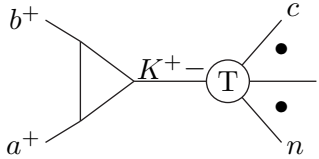
\begin{figure}[H]
  \def\rpt{20}\def\lpt{-15}\def\lht{15}\def\llb{-48}
  \begin{center}\begin{picture}(130,64)
      \SetOffset(65,32)
      \Line(\rpt,0)(\rpt,0)
      \Line(\rpt,0)(53,0)
      \Line(\rpt,0)(40,23)
      \Line(\rpt,0)(40,-23)
      \Line(\lpt,0)(\rpt,0)
      \Line(\lpt,0)(-35,-\lht)
      \Line(-35,-\lht)(-35,\lht)
      \Line(-35,\lht)(\lpt,0)
      \Line(-35,\lht)(\llb,23)
      \Line(-35,-\lht)(\llb,-23)
      \BCirc(\rpt,0){8}
      \Text(\rpt,0)[cc]{T}
      \Text(42,-25)[lc]{$n$}
      \Text(42,27)[lc]{$c$}
      \Text(12,1)[br]{$-$}
      \Text(-14,1)[bl]{$K^+$}
      \Text(-50,-23)[rc]{$a^+$}
      \Text(-50,24)[rc]{$b^+$}
      \Text(40,-10)[c]{\small$\bullet $}
      \Text(40,12)[c]{\small $\bullet $}       
    \end{picture}\end{center}
  \caption{Diagram contributing to the double-pole $\langle a\:b
    \rangle^{-2}$. The right-hand side is the appropriate tree-level
    amplitude.}
  \label{fig:doublepole}

\end{figure}

The factorisation arguments used above have all implicitly involved
real momenta. There is considerably more information available if we
consider complex momenta, but the factorisation properties of the
amplitudes are not so well understood. Also, double complex poles arise
in some amplitudes, for example in amplitudes with a single negative
helicity leg, both in Yang--Mills theory~\cite{Bern:2005hs} and
gravity~\cite{Dunbar:2010xk,Brandhuber:2007up}.  These double poles
are understood to arise from diagrams of the form illustrated in
\figref{fig:doublepole}.

Here there is one ${\spa a.b}^{-1}$ factor from the loop integral
associated with the all-plus triangle and a second from the propagator
joining the triangle to the tree amplitude. If the tree is
non-vanishing these give rise to contributions of the form
$$
{1\over \spa{a}.{b}^2}\times \cdots \equn
$$
which are double poles, but only in complex momentum space.
Explicitly, the three-point 1-loop all-plus amplitudes for Yang--Mills
theory and gravity are proportional to~\cite{Bern:2005hs,
  Dunbar:2010xk, Brandhuber:2007up}
\begin{equation}
  A^{(1){\text{Y--M}}}_3 =
  {{\spb a.b}{\spb a.K}{\spb K.b}\over s_{ab}}, \qquad
  M^{(1)\text{grav}}_3 =
  {{\spb a.b}^2{\spb a.K}^2{\spb K.b}^2\over s_{ab}},
\end{equation}
respectively, which vanish in the ${\spb a.b} \to 0$ limit and diverge
in the ${\spa a.b} \to 0$ limit.

These complex double poles are not present in all amplitudes. For
example, in the case of five-point MHV amplitudes the tree amplitudes
in \figref{fig:doublepole} have only one positive helicity leg and so
vanish. We can therefore insist that there are no higher-order complex
poles in $\spa a.b$ in the five-point MHV amplitudes we consider.
Similar arguments preclude any other complex higher-order poles in these
amplitudes.

\section{Examples of Spurious  and Multiple Singularities}
In this section we illustrate the types of unphysical singularities
that arise when we perform a reduction procedure and describe how they
cancel in the full amplitude.
   
As an example involving spurious singularities we consider one of the
six-point NMHV amplitudes in Yang--Mills.  (As usual we organise the
amplitudes according to external helicity.  Amplitudes with exactly
two-negative helicity legs are termed MHV, while those with exactly
three negative helicity legs are next-to-MHV, or NMHV.)  At six points
there are three independent colour-ordered NMHV amplitudes: $A_6
(1^-,2^-,3^-,4^+,5^+,6^+)$, $A_6 (1^-,2^-,3^+,4^-,5^+,6^+)$ and $A_6
(1^-,2^+,3^-,4^+,5^-,6^+)$.  Additionally, we organise according to
the matter content circulating in the loop, it being most convenient
to take a supersymmetric decomposition and consider three components
corresponding to a $\NeqFour$ multiplet, a $\NeqOne$ matter multiplet,
and a scalar particle circulating in the loop.  (See
ref.~\cite{Dunbar:2008zz} for an overview of this amplitude.)
The amplitude $A_6^{\Neq1,\text{matter}}( 1^-,2^-,3^-,4^+,5^+,6^+)$ is
given in ref.~\cite{Bidder:2004tx} as
\begin{equation}
  \eqalign{ A_6^{\Neq1,\text{matter}}( & 1^-,2^-,3^-,4^+,5^+,6^+) = {A^{\tree}\over 2} \bigl(I_2(s_{61}) + I_2(s_{34})\bigr) \cr 
    & - {i\over 2} \Biggl[ c_1 { L_0 [t_{345}/ s_{61} ] \over s_{61} } +c_2{
      L_0 [ t_{234} /s_{34} ] \over s_{34}} +c_3{ L_0 [ t_{234}/s_{61} ]
      \over s_{61}} +c_4{ L_0 [ t_{345} /s_{34} ] \over s_{34} } 
    \Biggr] \cr}
  \label{specificform}
\end{equation}
where the $c_i$ are
\begin{equation}
  \eqalign{ c_1 =& {  [6 |  K  | 3 \ra ^2[ 6 | [ k_2, K] K | 3 \ra 
      \over
      [ 2 |  K | 5 \ra \spb6.1\spb1.2\spa3.4\spa4.5 K^2 } \,, 
    \hskip 1.0 truecm K=K_{345} ;  \cr 
    c_2 =& { [ 4 | 
      K  | 1 \ra ^2[ 4 |  [K,k_2] K | 1 \ra 
      \over [ 2 |  K | 5 \ra
      \spb2.3\spb3.4\spa5.6\spa6.1 K^2 } \,, \hskip 1.0 truecm
    K=K_{234}  ; \cr 
    c_3 =& { [ 4 |  K | 1 \ra^2
      [ 4 |  [ k_5,K] K | 1 \ra 
      \over [ 2 |  K  | 5 \ra
      \spb2.3\spb3.4\spa5.6\spa6.1 K^2 } \,, \hskip 1.0 truecm K=K_{234} ; \cr
    c_4 =& {  [ 6 |  K  | 3 \ra ^2
      [ 6 | [ K, k_5]  K | 3 \ra \over  
      [ 2 |  K  | 5 \ra \spb6.1\spb1.2\spa3.4\spa4.5 K^2 } \,, \hskip 1.0
    truecm K=K_{345}. \cr}
  \label{SixPointNMHV}
\end{equation}
We have omitted an overall dimensional regularisation factor
$(\mu^2)^{\eps} c_\Gamma$ from the amplitude, where
\begin{equation}
  c_\Gamma\ =\ {r_\Gamma \over (4 \pi)^{2-\e}} \ , \qquad
  r_\Gamma \ =\
  {\Gamma(1+\e)\Gamma^2(1-\e)\over\Gamma(1-2\e)}.
  \label{cGamma}
\end{equation}

The function $L_0$ is defined by
$$
L_0[r] = { \ln(r) \over 1-r } \equn
$$
so that
$$
{ L_0[s/s'] \over s'} = {1\over s-s'} \left( I_2(s)-I_2(s') \right) 
\equn
$$ 
where  the scalar
bubble function is
$$
I_2(s) = \frac{1}{\eps} +2 -\ln(-s) +O(\eps). \equn
$$
The coefficient of $I_2(s_{34})$ is thus
\begin{equation}
  \eqalign{
    \frac{1}{2} A^{\tree} 
    &-\frac{i}{2} {1 \over t_{234}-s_{34} }  { [ 4 | 
      K_{234}  | 1 \ra ^2[ 4 |  [ K_{234} , k_2]  K_{234} | 1 \ra
      \over [ 2 |  K_{234} | 5 \ra
      \spb2.3\spb3.4\spa5.6\spa6.1 t_{234} } 
    \cr
    &-\frac{i}{2} {1 \over t_{345}-s_{34} } 
    {  [ 6 |  K_{345}  | 3 \ra ^2
      [ 6 | [ K_{345}, k_5]  K_{345} | 3 \ra \over  
      [ 2 |  K_{345}  | 5 \ra \spb6.1\spb1.2\spa3.4\spa4.5 t_{345} } 
    \cr}
\end{equation}
This coefficient contains a variety of singularities.  There are
physical singularities of the forms
$$
{ 1\over K^2} , \quad {1 \over \spa{a}.b} , \quad {1\over\spb{a}.{b} }
.  \equn
$$  There are also 
spurious singularities:
$$
{1 \over (t_{234}-s_{34}) } , \quad {1 \over (t_{345}-s_{34}) } ,
\quad {1\over [2|K_{34}|5\ra} \quad \equn
$$ 
which do not correspond to any physical singularity so must vanish in
the entire amplitude.

The first of these, $(t_{234}-s_{34})^{-1} $
arises as a Gram determinant in the reduction of the
two-mass tensor triangle integral with massless leg $k_2$ and a
massive leg $k_3+k_4$.   
This becomes a singularity when $k_2\cdot (k_3+k_4)=0$.   For real
momenta this occurs, provided $k_3$ and $k_4$ have opposite energy in
a two-parameter subspace which may be characterised by 
$k_2 =\alpha (\lambda_3+e^{i\theta}\lambda_4)(\bar\lambda_3-e^{-i\theta}\bar\lambda_4)$.
At this singularity $s_{23}-t_{234}
\longrightarrow 0$ so that $\ln(-s_{23}) \longrightarrow
\ln(-t_{234})$.  The singularity is present in the coefficients $c_2$
of \eqref{SixPointNMHV} which contributes to the coefficient 
of both $I_2(s_{34})$ and
$I_2(t_{234})$. It is the cancellation between the two
contributions when the integral functions degenerate into each other
that leaves the full amplitude finite.  The form of the
amplitude~\eqref{specificform} makes this simple to see since, as $r
\longrightarrow1$,
$$
L_0[r]={ \ln(r)\over (1-r) } \longrightarrow -1 -\frac{1}{2}(1-r)
+O\left((1-r)^2\right) \equn
$$
which is finite.

The final singularity in this amplitude, $[2|K_{34}|5\ra^{-1}$, occurs when $K_{34}$ is
co-planar with $k_2$ and $k_5$, i.e. $K_{34}=\alpha k_2 +\beta k_5$.
This singularity corresponds to the Gram determinant of a
two-mass-easy scalar box with massless legs $k_2$ and $k_5$ together
with massive legs $K_{34}$ and $K_{61}$.  At this point,
$$
t_{234}t_{345} -s_{34}s_{61} \longrightarrow 0 \equn
$$
or equivalently
$$
{ s_{34} \over t_{234} } \longrightarrow{ t_{561} \over s_{61} } \equn
$$
so the logarithms in $L_0[ t_{234}/s_{34} ]$ and
$L_0[t_{561}/s_{61}]$ may cancel.  In this case we are seeing a
cancellation between all four of the bubble integral functions
$$
c_1 I_2(s_{34})+c_2I_2(t_{234}) +c_3I_2(s_{61}) +c_4I_2(t_{345})
\longrightarrow 0. \equn
$$

The other type of singularity we consider are those that appear at the
same phase-space points as the physical singularities but are of
higher order.  These do not correspond to any singularity arising in
any Feynman diagram.  To distinguish this type of singularity from the
spurious singularities discussed above, we refer to them as
``higher-order physical'' poles.

\begin{figure}[H]
  \begin{center}
    {
      \begin{picture}(250,144)(0,0) \SetOffset(125,72)
        \Line(-40,0)(0,40) \Line(0,40)(40,0) \Line(40,0)(0,-40)
        \Line(0,-40)(-40,0) \Line(0,40)(0,60) \Line(0,-40)(0,-60)
        \Text(0,62)[bc]{$b^+$} \Text(0,-62)[tc]{$a^+$}
        \Line(40,0)(60,0) \Line(40,0)(55,15) \Line(40,0)(55,-15)
        \Line(-40,0)(-60,0) \Line(-40,0)(-55,15) \Line(-40,0)(-55,-15)
        \Text(62,1)[lc]{$m_2^-$} \Text(-62,1)[rc]{$m_1^-$}
        \Vertex(53,8){0.5} \Vertex(54,4.5){0.5} \Vertex(53,-8){0.5}
        \Vertex(54,-4.5){0.5} \Vertex(-53,8){0.5}
        \Vertex(-54,4.5){0.5} \Vertex(-53,-8){0.5}
        \Vertex(-54,-4.5){0.5} \Text(-75,0)[rc]{$\displaystyle M
          \left\{\vphantom{\frac12}\right.$}
        \Text(75,0)[lc]{$\displaystyle
          \left.\vphantom{\frac12}\right\}N$}
      \end{picture}
    }
    \\
    \caption{ The two-mass-easy box functions appearing in the MHV one-loop
      amplitudes. The cases where $M$ or $N$ contain a single negative helicity leg and the integral reduces to the one-mass box are
included. \label{BoxFigure} }
  \end{center}
\end{figure}
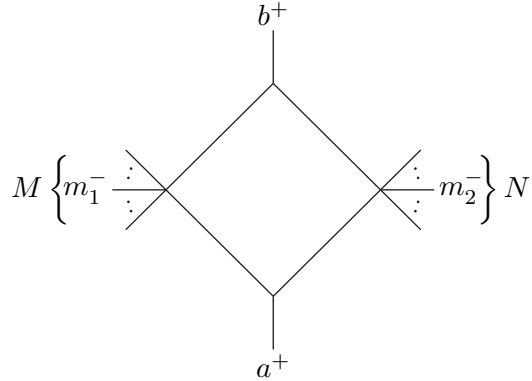

To illustrate how these arise and cancel we can consider one-loop
$n$-point MHV Yang--Mills amplitudes. 
The boxes depicted in \figref{BoxFigure} have non-vanishing coefficients. Denoting the two negative helicities as $m_1$ and
$m_2$ and considering the box with two massless legs $a$ and $b$, the
leading colour coefficients of the box integrals
are~\cite{Bern:1994zx,Bern:1994cg,Bedford:2004nh,BBDP}
$$
\eqalign{ a^{ \NeqFour} =& (st-M_2^2M_4^2) \Atree \,, \cr 
a^{ \NeqOne}=& (st-M_2^2M_4^2) \Atree \times { B_{ab}^{m_1m_2} }\,,
  \cr a^{ \NeqZero}=& (st-M_2^2M_4^2) \Atree \times { (
    B_{ab}^{m_1m_2} ) ^2 }\,, \cr} \equn$$ where
$$
B_{ab}^{m_1m_2} = { \spa{a}.{m_1} \spa{a}.{m_2} \spa{b}.{m_1}
  \spa{b}.{m_2} \over \spa{a}.{b}^2 \spa{m_1}.{m_2}^2 } \; , \equn
$$
and
$$
s= (k_a+K_M)^2 ,\;\;\; M_2^2=K_M^2,\;\;\; t=(K_M+k_b)^2 ,\;\;\;
M_4^2=K_N^2. \equn$$

These coefficients contain $\spa{a}.b^{-n}$ singularities. Near these
singularities the box integral functions can be expanded as
$$  { (st-M_2^2M_4^2)\over 2 r_\Gamma } I_4
=\sum_{s_i=s,t,M_2^2,M_4^2} \Biggl( {\pm 1\over
  \epsilon^2}(-s_i)^{-\epsilon} + s_{ab} P_i \ln(-s_i) \Biggr) +
s_{ab}^2 P_R, \equn
$$
where the $P_i$ and $P_R$ are rational functions of the momentum
invariants and specifically are polynomials in $s_{ab}$.  Thus, as we
approach the singularity these box contributions {\it degenerate}, the
first term of the sum cancels with the one- and two-mass triangle contributions as
discussed in more detail in the next section, leaving logarithmic and
rational {\it descendants} of the box. The logarithmic descendants
combine with the logarithms in the bubble contributions to cancel the
higher-order physical poles in their coefficients. Similarly, any
higher-order physical poles in the rational descendants cancel against
the rational piece of the amplitude, $R_n$.

\section{One-Loop Gravity Amplitudes}

A one-loop graviton scattering amplitude can receive contributions
from a range of particle types circulating in the loop. It is
convenient to perform a supersymmetric decomposition and compute the
contributions from entire matter supermultiplets circulating in the
loop.  The specific particle contributions are then simply obtained as
linear combinations of the supersymmetric contributions:
$$
\eqalign{
  M^{[1/2]} &= M^{\Neq1}- 2M^{[0]} \cr M^{[1]} &=
  M^{\Neq4}-4M^{\Neq1}+2M^{[0]} \cr M^{[3/2]} &=
  M^{\Neq6}-6M^{\Neq4}+9M^{\Neq1}-2M^{[0]} \cr M^{[2]} &=
  M^{\Neq8}-8M^{\Neq6}+20M^{\Neq4}-16M^{\Neq1}+2M^{[0]} \cr }
\equn\label{InverseDecomp}
$$
where the superscript ${[s]}$ denotes a particle of spin $s$
circulating in the loop. We will sometimes refer to the contribution
from a real scalar, $M^{[0]}$, as $M^{\NeqZeroF}$.
 
In general, amplitudes with greater supersymmetry have simpler
structure and are easier to compute. The $\NeqEight$ one-loop
amplitudes have a particularly simple form consisting only of
box-functions~\cite{BjerrumBohr:2006yw, MaxCalcsB,MaxCalcsC, BjerrumBohr:2008ji}, a feature
shared with $\NeqFour$ Yang--Mills and related to the possible
finiteness of maximal supergravity.
This simplicity arises from cancellations between diagrams that reduce
the effective degree, $\deff$, of the loop momentum polynomial $P(\ell)$
in \eqref{BasicLoopEq}. The traditional expectation for $m$-point
supergravity amplitudes is that cancellation between particle types
within a supermultiplet reduces $\deff$ from $2m$ to $2m-r$, where $r$
depends upon the degree of supersymmetry.

For $\Neq8$ supergravity, that $r=8$ is manifest term by term within
the ``string-based rules'' method~\cite{GravityStringBasedA,
  GravityStringBasedB}.  However the no-triangle hypothesis indicates
that further cancellations arise, resulting in $\deff=m-4$.  This
suggests a degree of $m + 4$ (rather than $2m$) for pure gravity,
reduced by a further $8$ for $\Neq8$ supersymmetry.
For $\NeqSix$ supergravity $\deff=m-3$ and for $\NeqFour$ supergravity
$\deff=m$~\cite{Bern:2007xj,Dunbar:2010fy}. In the former case the amplitudes have
box and triangle contributions only, while in the latter we also have
bubbles and purely rational terms. The rational terms are not four
dimensional 
cut-constructible.

It is conventional to express the $\NeqSix$ and $\NeqFour$ amplitudes
in terms of a basis involving truncated box functions and three-mass
triangle functions. While this packaging can be motivated by IR
arguments, singularity arguments provide an alternative explanation.
In a basis involving box and triangle functions the $\Neq6$ one-loop
amplitudes take the form
$$
M_n^{\oneloop,\NeqSix}=\sum_{i\in \cal C}\, a_i\, I_4^{i} +\sum_{j\in
  \cal D}\, b_{j}\, I_3^{j} \equn\label{DecompBasisN6}
$$ 

For the MHV configuration, the sum of boxes is fairly restrictive,
consisting only of boxes with two massive and two massless legs where
the massive legs are non-adjacent.  The two negative helicities must
lie on the opposite clusters of legs denoted $M$ and $N$ in \figref{BoxFigure}.  (We also
include in the sum the degenerate one-mass case where one of the
negative helicity legs is on its own.)
The box-coefficient is given by
\begin{equation}
  {(-1)^n\over 8} \spa1.2^8  \left( -{ \spa{1}.{a}\spa{2}.{a}\spa{1}.{b}\spa{2}.{b} \over
      \spa{a}.b^2 \spa{1}.{2}^2 } \right) h(a, M, b) h(b, N, a)
  \tr^2[a\, M\, b\, N]\
  \label{MHVN6}
\end{equation}
where $\tr[a\, M\, b\, N]=\tr[\slashed {k}_a\, \slashed {k}_M\, \slashed {k}_b\, \slashed {k}_N]$.
The $h(a, M, b)$ are the ``half-soft'' functions of
ref.~\cite{MaxCalcsB},
$$
\eqalign{ h(a,\{1,2,\ldots,n\},b) &\equiv \frac{\spb1.2}{\spa1.2} { [3
    | {K_{12}} |a \ra [4 | {K_{123}} |a \ra \cdots [n| {K_{1\cdots
        n-1}} |a \ra \over \spa2.3\spa3.4 \cdots \spa{n-1,}.{n} \,
    \spa{a}.1 \spa{a}.2\spa{a}.3 \cdots \spa{a}.{n} \, \spa1.{b}
    \spa{n}.{b} } \cr &\hskip1cm + \Perm(2,3,\ldots,n). \cr}
\equn\label{NonRecursiveH}
$$
 As we can see
this coefficient contains a higher-order physical singularity,
$\spa{a}.{b}^{-2}$.  Since this is unphysical it must cancel within
the amplitude. Using the explicit forms of the one- and two-mass
triangle integral functions and their coefficients we can repackage
the expansion:
\begin{equation}
  \begin{split}
    M_n^{\oneloop,\NeqSix} &= \sum_{i\in \cal C}\, a_i\, I_4^{i}
    +\sum_{j\in \cal D}\,
    b_{j}\, I_3^{j} \\
    &= \sum_{i\in \cal C}\, a_i\, \left( I_4^{i} +\sum_{j\in
        \mathcal{D}|_{\text{$1,2$-mass}}} \tilde b_{ij} I_3^{j}
    \right)+\sum_{j\in \mathcal{D}|_{\text{$3$-mass}}} b_j  I_3^{j} \\
    &= \sum_{i\in \cal C}\, a_i\, I_4^{i,\trunc} +\sum_{j\in
      \mathcal{D}|_{\text{$3$-mass}}} b_j I_3^{j} ,
  \end{split}
  \label{DecompBasisN6trunc}
\end{equation}
where $I_4^{i,\trunc}$ is of order $s_{ab}$ near $s_{ab}=0$. The
explicit form of the one and ``two-mass easy'' truncated boxes is
given in appendix~B.

This result is well-known and is normally interpreted as a
cancellation of spurious IR singularities~\cite{Bern:1994cg,BBCFsusyone,BBDP} .  Here we wish to note that
the truncated boxes can be obtained by requiring the cancellation of
higher-order (non-IR) singularities and as such, this is an example
where the box-coefficients contain sufficient information to
reconstruct the entire amplitude from singularity considerations.

\section{Four-Point Gravity Amplitudes}

In this section we discuss the simple example of four-point gravity
MHV amplitudes, where the entire
amplitude can be constructed from the box-coefficients.
The MHV amplitude
$M_4(1^-,2^-,3^+,4^+)$ contains all three four-point boxes for the
$\NeqEight$ multiplet but for ${\cal N}<8$ the $s=s_{12}$ unitarity
cut vanishes identically and we deduce the amplitude has the form
$$
a_4 I^{\trunc}_4(s_{13},s_{23})+ c_1 I_2(s_{23})+c_2 I_2(s_{13}) +R
\equn
$$
with only the box with ordering of legs $1324$ appearing.  As
discussed previously, we have combined the box with  triangle
contributions $I_3(s_{23})$  and $I_3(s_{13})$ into a truncated box functions. 
The
coefficient of the box could easily be derived using quadruple cuts~\cite{BrittoUnitarity}
and is
$$
a_4 =
\left( {s_{12}s_{23} \spa1.2^4 \over \spa1.2\spa2.3\spa3.4\spa4.1}
\right)^2 \times \left( { s_{23}s_{13} \over s_{12}^2 } \right)^A
\equn
$$
with $A=0,1,2,4$ for the $\NeqEight,4,1$ and $0$ multiplets
respectively.  These amplitudes have an increasingly high order
singularity when $s_{12}\to0$ if we allow ourselves to consider
complex momenta where $\spb1.2\to 0$ but $\spa1.2\neq 0$.

Setting $s=s_{12}$, $t=s_{23}$ and $u=s_{13}$, and suppressing a factor of
$ic_\Gamma$, the amplitude in the
$\Neq0$ case (with a real scalar in the loop) takes the form
\begin{multline}
\left( {st \spa1.2^4 \over \spa1.2\spa2.3\spa3.4\spa4.1} \right)^2
\Biggl( -{ t^3u^3\over s^8 } \left( \ln^2(t/u)+\pi^2\right) +
\\
 { a(t,u) } \left( -\frac{1}{\epsilon} -2+\ln(-t) \right)+ {
    a'(t,u) }\left( -\frac{1}{\epsilon} -2+\ln(-u) \right) 
+b(t,u) \Biggr) 
\label{fourptansatzi}
\end{multline}
where, from the symmetry in the amplitude, $a'(t,u)=a(u,t)$ and
$b(u,t)$ must be symmetric in $(u,t)$. 
The $\eps^{-1}$ infra-red singularity vanishes in this
amplitude~\cite{DunNorB} so $a'(t,u)=-a(t,u)$ and the amplitude takes
the form
$$
\left( {st \spa1.2^4 \over \spa1.2\spa2.3\spa3.4\spa4.1} \right)^2
\Biggl( -{ t^3u^3\over s^8 } \left( \ln^2(t/u)+\pi^2\right) +
 { a(t,u) } \ln(t/u)
+b(t,u) \Biggr) 
\equn
\label{fourptansatz}
$$
with $a(t,u)=-a(u,t)$. 

 We are interested in the
behaviour of the amplitude as $\spb1.2\to 0$. As $s_{12}\to 0$
momentum conservation requires $s_{34}\to 0$. If we utilise a shift to
approach the singular point~\cite{BCFW},
$$
\bar\lambda_1\to \bar\lambda_1 -z \bar\lambda_3, \qquad \lambda_3\to
\lambda_3 +z\lambda_1, \equn
$$
then $\spb3.4$ is unshifted and $\spa3.4$ must vanish along with
$\spb1.2$. The first factor in \eqref{fourptansatz} is finite in this
limit and the leading singularity is order $\spb1.2^{-8}$.

If we expand about $s=0$, using $s+u+t=0$ then
$$
\ln(t/u)^2 = -\pi^2 +2i\pi {s \over u} +\cdots .  \equn
$$
The first term can be cancelled by $b(u,t)$, while the second must be
cancelled by the bubble contributions.

Using the expansion,
$$
\ln(t/u)= i\pi + {s\over u } +\cdots \equn
$$
we see that $a(t,u)$ must contain a factor of $s^{-7}$ and must be
anti-symmetric in $t$ and $u$.  Since it is rational it must therefore
contain a factor of $(t-u)$.  We can thus take
$$
a(t,u)= { (t-u)( \alpha u^4 +\beta u^3t +\gamma u^2t^2 +\beta ut^3
  +\alpha t^4 ) \over s^7} \equn
$$
Requiring the cancellation of the $s^{-7}$ singularity imposes one
constraint on the parameters:
$$
2\alpha-2\beta+\gamma-1=0.  \equn
$$
If this constraint is satisfied, the cancellation of the $i\pi s^{-6}$
singularity is automatic. There are four further constraints arising
from demanding the cancellation of the $i\pi s^{-5}$ through $i\pi
s^{-2}$ singularities. Fortunately only two of these constraints are
independent and we have a well posed problem.  Solving the system of
constraints gives
$$
\alpha=\frac{1}{30}, \qquad \beta =\frac{9}{30}, \qquad \gamma
=\frac{46}{30}.  \equn
$$

Cancellations involving $a(t,u)$ remove all the higher-order poles that
have an $i\pi$ factor, but singularities with no $i\pi$ factor remain.
We therefore add the most general form for $b(t,u)$ consistent with
the symmetries discussed above:
$$
b(t,u) = { \delta u^4 +\eta u^3 t + \zeta
  u^2 t^3 + \eta u t^3 +\delta t^4 \over s^6 } \equn
$$
Requiring that the $s^{-6}$ through $s^{-2}$ singularities cancel
again gives just three independent constraints. Solving these gives
$$
\delta=\frac{2}{180}, \qquad \eta=\frac{23}{180}, \qquad
\zeta=\frac{222}{180}.  \equn
$$
The full one-loop amplitude is then 
\begin{multline}
  M_4^{[0]}(1^-,2^-,3^+,4^+) = ic_\Gamma\left( {st \spa1.2^4 \over
      \spa1.2\spa2.3\spa3.4\spa4.1} \right)^2 \times \Biggl( -{
    t^3u^3\over s^8 } ( \ln^2(t/u)+\pi^2) +\\ {( t-u)(
    t^4+9ut^3+46u^2t^2 +9u^3t+u^4) \over 30 s^7 } \ln({t/ u}) + { 2
    t^4+23ut^3+222u^2t^2+23u^3t+2u^4 \over 180 s^6 } \Biggr) \
\end{multline}
This exactly matches the amplitude previously calculated using the
string-based rules for gravity~\cite{GravityStringBasedA,
  GravityStringBasedB}.  The corresponding analyses for the $\NeqOne$
and $\NeqFour$ amplitudes are progressively simpler and again
higher-order pole constraints are sufficient to determine the
amplitudes completely.

\section{ $\NeqFour$ five points one-loop amplitudes}

The $n$-graviton MHV amplitude for a $\NeqFour$ matter multiplet is
\begin{multline}
  M_n^{\NeqFour} (1^-, 2^-, 3^+, \ldots, n^+) = \\
  \quad {(-1)^n\over 8} \, \spa1.2^8 \sum_{2 < a < b \leq n \atop 1\in
    M, 2 \in N} \left( -{ \spa{1}.{a}\spa{2}.{a}\spa{1}.{b}\spa{2}.{b}
      \over \spa{a}.b^2 \spa{1}.{2}^2 } \right)^2 h(a, M, b) h(b, N,
  a)
  \tr^2[a\, M\, b\, N]\, I_4^{aMbN, \trunc} \ \\
  \quad\quad+\sum_{1\in A,2\in B} c_2(A;B) I_2 (K_A^2) +R_n,\hfill
  \label{MHVboxes}\end{multline}
where the sets $A$ and $B$, contain at least one positive helicity
leg.  As discussed previously, we have combined the box and triangle
contributions to the amplitude to leave a sum over the box
coefficients multiplied by truncated box functions. For the five-point amplitude the box-coefficients are all
equivalent up to relabelling and, for example,  the coefficient of 
$I_4^{3\{1\}4\{25\}}$  reduces to,
$$
-\frac{1}{2} \spa1.2^4 \times {1\over \spa3.4^4 } \times { \spb2.5 s_{14}^2s_{13}^2
  \spa2.3\spa2.4\over\spa2.5\spa3.5\spa4.5} \equn
$$
which clearly has a higher-order pole: $\spa3.4^{-4}$ .
The limit $\spa3.4\to 0$ corresponds to $u\to 0$ in Mandelstam notation. Close to $u=0$ we can
expand the truncated box functions:
\begin{equation}
  \begin{split}
    \eqalign{ I_4^{\trunc}(s,t,u,m^2) = &+{2u \over (st)^2} \left(
        m^2\ln(-m^2) - s \ln(-s) - t \ln(-t)\right) \\ &-{ u^2 \over (
        st)^3} \left( m^4\ln(-m^2) - s^2 \ln(-s) - t^2 \ln(-t) +(st)
      \right) +{O}(u^3) \cr}
  \end{split}
\end{equation}
The ${O}(u^3)$ terms combine with the coefficient to yield a
$\spa3.4^{-1}$ singularity and so only contribute to physical
singularities.

The logarithms arising from the expansion of $I_4^{\trunc}$ around
$u=0$ can be combined with those from the bubble contributions. The
bubble contributions are presented in general form in
appendix~\ref{BubblesAppendix}.  
The resulting coefficients of the logarithms have only simple poles in $u$
and therefore in the $u\to 0$ limit only produce logarithmic
contributions to physical singularities/factorisations.

The rational term or {\it descendent} arising from the expansion of
$I_4^{\trunc}$ contains a higher-order pole which must be cancelled by
$R_5$.  We therefore introduce
$$
\eqalign{ R_5^a =& - { \spa1.2^4 \over 2} {\spb3.4^2\over \spa3.4^2}
  {\spb2.5 \spa2.3\spa2.4\over \spa2.5 \spa3.5\spa4.5 }
  +\Perm(\{1,2\},\{3,4,5\}) \cr} \equn
$$
where $\Perm(\{x_i\},\{y_i\},\dots)$ denotes a sum over the
\emph{independent} permutations of the $x_i$ and $y_i$,
and express the full rational term as
$$
R_5 =R_5^a +R_5^b.  \equn
$$
As $R_5^a$ cancels all of the higher-order poles descending from the
boxes, $R_5^b$ contains only physical singularities.  Assuming that
any non-standard complex factorisations are restricted to the
$\spa{c}.{d}^{-1}$ pieces of collinear limits involving two positive
helicity legs, we find that the only singularities that $R_5^b$ can have take the
form $\spa{c}.{d}^{-1}$ where $c$ and $d$ denote positive helicity
legs. There is only one combination of spinor products involving only
these poles that has the correct spinor and momentum weights.  The
normalisation of this term can be determined by evaluating real
collinear limits, yielding
$$
R_5^b = -\spa1.2^4 {\spb3.4\spb3.5\spb4.5\over \spa3.4\spa3.5\spa4.5 }
\; . 
\equn
$$
Explicit computation using string based-rules numerically verifies $R_5$~\cite{Dunbar:2010fy} which  has also been deduced using
colour--kinematics duality applied to gravity~\cite{Bern:2011rj}.

\section{$\NeqOne$ five point one-loop amplitude}

The pole structure for the five-point $\Neq1$ amplitude is much richer
than the $\Neq4$ case and contains both higher-order physical poles
and spurious singularities. Starting with the contributions of the
boxes to the amplitude, the truncated box integral functions have
coefficients:
$$
a^{\Neq1}[\{a^-,e^+\},c^+,b^-,d^+]= {\spa{a}.{b}^2 \spa{a}.{c}^2
  \spa{a}.{d}^2 \spa{b}.{c}\spa{b}.{d} {\spb a.e} s_{bd}^2 s_{bc}^2
  \over 2\spa{a}.{e}\spa{c}.{d}^6 \spa{c}.{e} \spa{d}.{e} }.  \equn
$$  
Close to the $\spa{c}.{d}=0$ pole the box contributions generate
logarithms with coefficients containing poles up to $\spa{c}.{d}^{-5}$ and
rational descendants with coefficients containing poles up to $\spa{c}.{d}^{-4}$.
The higher-order poles in these rational descendants must be cancelled by
the rational piece of the amplitude.  The rational terms needed to
cancel the $\spa{c}.{d}^{-4}$ and $\spa{c}.{d}^{-3}$ poles in the
rational descendants are easily obtained from the expansion of the
truncated box integral function. For each box we introduce a rational
term,
\begin{align}
R^{\spa{c}.{d}^{-4},\spa{c}.{d}^{-3}}_{\{a^-,e^+\},c^+,b^-,d^+} &=
{\spa{a}.{b}^2 \spa{a}.{c}^2 \spa{a}.{d}^2 \spa{b}.{c}\spa{b}.{d}
  {\spb a.e} s_{bd}^2 s_{bc}^2 \over 2\spa{a}.{e}\spa{c}.{d}^6
  \spa{c}.{e} \spa{d}.{e} } \left( {s_{cd}^2 \over s_{bc}^2 s_{bd}^2}
  - {s_{cd}^3 s_{ae} \over 3 s_{bc}^3 s_{bd}^3 }\right)  \notag
\\
&={\spa{a}.{b}^2 \spa{a}.{c}^2 \spa{a}.{d}^2 \spa{b}.{c}\spa{b}.{d}
  {\spb a.e} \over 2\spa{a}.{e}
  \spa{c}.{e} \spa{d}.{e} } \left( {\spb{c}.{d}^2\over \spa{c}.{d}^4 }
  + {\spb{c}.{d}^3 s_{ae} \over 3 \spa{c}.{d}^3s_{bc} s_{bd} }\right).
\end{align}
These rational pieces do not contain additional higher-order poles, but do
not cancel the $\spa{c}.{d}^{-2}$ poles in the rational descendants.
If the expansion of the truncated box integral function is taken a
stage further, the rational term that is naively generated contains
double poles in $s_{bc}$ and $s_{bd}$. To deal with the
$\spa{c}.{d}^{-2}$ poles we must also consider the bubble
contributions to the amplitude.

The bubble coefficients are readily evaluated using the canonical
basis procedure ~\cite{UsUnitarity}. The full $\NeqOne$ bubble
coefficient is given in appendix \ref{BubblesAppendix}. These
coefficients contain a number of higher-order physical and spurious poles:
$$
c^{\Neq1}[\{a^-,c^+\},\{b^-,d^+,e^+\}] \supset \{ \spa{c}.{d}^{-5},
\spa{c}.{e}^{-5}, \spa{d}.{e}^{-5}, (d\cdot K_{ac})^{-3}, (e\cdot
K_{ac})^{-3}\}.  \equn
$$
The $\spa{c}.{d}^{-5}$ type poles are precisely those needed to cancel
the logarithmic descendants of the boxes on $\spa{c}.{d} \to 0$ type
singularities. In fact, combining the logarithms from the bubbles with
those descending from the boxes gives logarithms with simple poles in
$\spa{c}.{d}$ as $\spa{c}.{d} \to 0$. Therefore there are no rational
terms descending from the logarithms as we approach this type of
singularity.

Each spurious pole occurs in two bubbles as $d\cdot K_{ac}=-d\cdot
K_{be}$ etc. As the logarithms themselves cannot contain a spurious
pole, the coefficients of each pair of logarithms must cancel to order
$(d\cdot K_{ac})^{-1}$, so as we approach $d\cdot K_{ac}=0$ the bubble
contributions combine to give,
\begin{equation}
  \begin{split}
    c^{\NeqOne} &\left(
      -\ln(-s_{ac})+\ln(-s_{be})\right) +{O}\left( (d\cdot
      K_{ac})^0\right) \\
    &=c^{\NeqOne}\ln\left(1+{d\cdot
        K_{ac}\over
        s_{ac}}\right) +{O}\left( (d\cdot K_{ac})^0\right)\\
    &=c^{\NeqOne}\left({d\cdot
        K_{ac}\over s_{ac}} -{1\over 2}{(d\cdot K_{ac})^2\over
        s_{ac}^2}\right) +{O}\left( (d\cdot K_{ac})^0\right)
  \end{split}
\end{equation}
The rational descendants of the logarithms on the spurious pole
contain both spurious and higher-order physical poles, all of which must
be cancelled by $R_5$. As the spurious poles do not appear in the box
coefficients, it is natural to remove them before we combine the box
and bubble induced rational pieces. The full bubble coefficient is
quite complicated, but the term containing the leading spurious pole
is much simpler:
$$
c^{\NeqOne}[\{a^-.c^+\},\{b^-,d^+,e^+\}]= {\spa{a}.{b}^2\spa{a}.{c}^2\spa{a}.{d}  
  \spa{b}.{d}^3  {\spb a.c}{\spb b.e} {\spb
    c.d}^3 {\spb d.e} \over 12 \spa{c}.{d}^2
  \spa{d}.{e}^2 (2d\cdot K_{ac})^3} +{O}\bigl( (d\cdot
K_{ac})^{-2}\bigr).  \equn
$$
The leading order descendant is simply this multiplied by $ {d\cdot
  K_{ac}/ s_{ac}} $. On the spurious pole $s_{ad}=-s_{cd}$ and
$s_{de}=-s_{bd}$, allowing us to construct a factor that has
adjustable sub-leading behaviour:
$$
\left(\alpha+(\alpha-1){s_{cd}\over s_{ad}} \right)\left(\gamma
  +(\gamma-1){s_{de}\over s_{bd}}\right) = 1+ {O}\left( d\cdot
  K_{ac}\right).  \equn
$$
For each bubble we introduce a rational term,
\begin{equation}
  \begin{split}
    R^{\rm spur}_{\{a^-.c^+\},\{b^-,d^+,e^+\}} &= {\spa{a}.{b}^2\spa{a}.{c}^2\spa{a}.{d}
      \spa{b}.{d}^3   \spb a.c\spb b.e {\spb
        c.d}^3  \spb d.e \over 24 \spa{c}.{d}^2
      \spa{d}.{e}^2
      (2d\cdot K_{ac})^2 s_{ac}} \times \\
    &\quad \left(\alpha_d+(\alpha_d-1){s_{cd}\over s_{ad}}
    \right)\left(\gamma_d +(\gamma_d-1){s_{de}\over s_{bd}}\right) +(d
    \leftrightarrow e).
  \end{split}
\end{equation}
The extra factor of $1/2$ arises as each descendant originated from a
pair of bubbles. By construction this cancels the leading spurious
pole for any values of $\alpha_d$, $\alpha_e$, $\gamma_d$ and
$\gamma_e$.  Additionally, this term is found to cancel the $(d\cdot
K_{ac})^{-1}$ poles if,
$$
\gamma_d = - 2-\alpha_e, \qquad \gamma_e = - 2-\alpha_d.  \equn
$$
Thus the spurious poles have been removed and we don't need to refer
to the bubble coefficient again. It is also worth noting that the only
double physical poles in $R^{\rm spur}$ are those that appear
explicitly in the expansion of the bubble coefficient.

The remaining higher-order poles involve factors of the form
$\spa{c}.{d}^{-2}$. These are present in the rational terms we have
already introduced and descend from the box contributions in the
$\spa{c}.{d}\to 0$ limit. Setting $\alpha_d=\alpha_e=0$ restricts the
higher-order poles to a small number of terms in $R^{\rm spur}$. Focussing
on terms containing $\spa{c}.{d}^{-2}$ specifically, it is possible to
rewrite the sum of these terms in a form that involves no spurious
poles and no other higher-order poles. The $\spa{c}.{d}^{-2}$ poles can
then be cancelled by introducing a rational term:
$$
R^{{\rm quad}:cd}=R^{\rm quad:X}+R^{\rm quad:Y}, \equn
$$
where
$$ 
R^{\rm quad:X}={ 5 \spa{a}.{b}^2\spa{a}.{c}^2\spa{a}.{d}^2 {\spb
    a.e}{\spb c.d}^4 \over
  24\spa{a}.{e}\spa{c}.{d}^2\spa{c}.{e}\spa{d}.{e}{\spb b.c}{\spb
    b.d}}+(a\leftrightarrow b), \equn
$$       
and
\begin{multline}
  R^{\rm quad:Y} = {\spa{a}.{b}^2\spa{b}.{c}{\spb c.d}\over 12\spa{c}.{d}^2\spa{c}.{e}{\spb a.c}{\spb b.c} }
  \Biggl\{ 
   -\spa{a}.{b} {\spb b.d}  {\spb c.e}^3  
   +\spa{a}.{d}{\spb c.d}{\spb c.e}^2{\spb d.e} 
  \\
  + \spa{a}.{c} \spa{a}.{d}{\spb a.e} {\spb c.d}^2\left( {
      \spa{a}.{b}\spa{a}.{d}{\spb a.c}+\spa{a}.{e}\spa{b}.{d}{\spb c.e} 
       \over 
      \spa{a}.{e}\spa{b}.{c} \spa{d}.{e} } \right)
  \Biggr\} +\Perm(\{ a, b\},\{ c, d\}).
\end{multline}
The $\spa{c}.{e}^{-2}$ and $\spa{d}.{e}^{-2}$ poles can similarly be
cancelled by introducing rational terms that are relabellings of
$R^{{\rm quad}:cd}$.

The full rational piece of the amplitude can now be written as,
$$
R_5 = \left(\sum_{{\rm boxes}:\,i}
  R^{\spa{c}.{d}^{-4},\spa{c}.{d}^{-3}}_i\right) +\left(\sum_{{\rm
      bubbles}:\,j} R^{\rm spur}_j \right) +R^{{\rm quad}:cd} +
R^{{\rm quad}:ce} + R^{{\rm quad}:de} +R_5^b, \equn
$$
where $R_5^b$ contains only simple physical poles.

To determine $R_5^b$ we apply complex factorisation constraints, again
assuming that any non-standard factorisations are restricted to the
$\spa{c}.{d}^{-1}$ pieces of limits involving two positive helicity
legs. Firstly: there should be no poles as either $\spa{a}.{b}\to 0$
or $[ab]\to 0$. The amplitude already satisfies these constraints.
Next, we expect no poles of the form $\spb{a}.{c}^{-1}$.  As such
poles are present in the rational terms we have already identified, we
introduce a further rational term to cancel them:
$$
R^{\rm sq}= { \spa{a}.{b}^2 {\spb c.d}^3 {\spb c.e}{\spb d.e}
  (s_{ce}s_{de} -s_{ae}s_{be}) \over 12 {\spb a.c}{\spb a.d}{\spb
    b.c}{\spb b.d} \spa{c}.{d}\spa{c}.{e}\spa{d}.{e} } + (\text{cyclic
  perms $\{c,d,e\}$}).  \equn
$$
Note that this term introduces no higher-order physical poles, spurious
poles, $\spa{a}.{b}$ poles or $\spb a.b$ poles, so it doesn't disrupt
any of the previous cancellations.
With the introduction of $R^{\rm sq}$ the $\spa{a}.{c}$ type poles in
the rational term have the correct coefficients to reproduce the
standard factorisations as $k_{a^-}\cdot k_{c^+} \to 0$ etc.

The only remaining poles involve terms of the form ${\spb c.d}^{-1}$
and $\spa{c}.{d}^{-1}$. We expect no factorisations to contribute to
the former type of pole and the amplitude has none. This leaves terms
involving $\spa{c}.{d}^{-1}$ type poles. The only term with the
correct spinor and momentum weights involving only these poles is,
$$
R^{\rm of} =-{1\over 4} \spa{a}.{b}^4{{\spb c.d}\over \spa{c}.{d}}
{{\spb c.e}\over \spa{c}.{e}} {{\spb d.e}\over \spa{d}.{e}}, \equn
$$
where the normalisation is fixed by examining the $k_{c^+}\cdot k_{d^+} \to
0$ collinear limit.

The full rational piece of the amplitude is thus,
$$
R_5= \Bigl(\sum_{{\rm boxes}:\,i}
R^{\spa{c}.{d}^{-4},\spa{c}.{d}^{-3}}_i\Bigr) +\Bigl(\sum_{{\rm
    bubbles}:\,j} R^{\rm spur}_j \Bigr) +R^{{\rm quad}:cd} + R^{{\rm
    quad}:ce} + R^{{\rm quad}:de} + R^{\rm sq} + R^{\rm of}.  \equn
$$
With this rational piece,  we have an ansatz for the amplitude which
a) is free of spurious poles, b)  has the correct symmetries, c) has the correct pole structure,
collinear and soft limits.  Since, any potential ambiguity  must
vanish in all limits, we expect this ansatz to be correct.

\section{Conclusions}

The absence of spurious singularities from complete one-loop
scattering amplitudes introduces a constraining web of relationships
between the rational functions arising when the amplitude is expanded
in terms of scalar one-loop integrals.  In particular, these
constraints involve the purely rational pieces of the amplitude as
well as the four-dimensional cut-constructible pieces. As the latter
are relatively easily determined from unitarity considerations, this web of constraints
readily provides information about the purely rational pieces.

In the simplest cases, for example the four-graviton amplitudes
considered in this article, these constraints determine the entire
one-loop amplitude starting from the coefficients of the box integral
functions. For more complicated examples, such as the five-graviton
MHV amplitudes, the web of constraints determine a significant portion
of the purely rational terms. The remainder has a relatively simple
form, which can be determined from the symmetries and the physical
factorisation properties of the amplitudes.  We have
successfully used this to reproduce the one-loop five-point $\Neq4$
supergravity amplitude, and obtain the previously-unknown one-loop
five-point amplitude for $\Neq1$ supergravity.

This research was supported by the STFC of the UK.

\appendix

\section{Bubbles in Supergravity MHV amplitudes}
\label{BubblesAppendix}
The bubble integral functions $I_2(P^2)$ will have vanishing
coefficients for the MHV amplitude unless the momenta $P$ (and hence
$-P$) contain exactly one negative helicity leg and at least one
positive helicity leg.  We can thus take $P$ of the form
$\{m_1^-,a_1^+,a_2^+,\cdots, a_{n_L}^+\}$ and the legs on the other
side to be $\{m_2^-,b_1^+,b_2^+,\cdots, b_{n_R}^+\}$.

Here we present the bubble coefficients for MHV amplitudes for the
$\NeqFour$ and $\NeqOne$ matter multiplets.  There are a variety of
techniques available to determine the bubble coefficient from the cut:
we will use the method of canonical forms~\cite{UsUnitarity}. The
$\NeqFour$ coefficients appear in ref.~\cite{Dunbar:2010fy}:
\begin{multline}
  \label{eq:bubcoeff1}
  c^{\NeqFour} [\{m_1,a_i\} ;\{m_2,b_i\} ] ={1\over 2} { \spa{m_1}.{m_2}^4
  }\sum_{
    P_L,P_R } C_{P_L} C_{P_R} \biggl( \\
  \sum_{x\neq a_1} D_x { \spa{m_2}.{a_1} \over \spa{b_1}.{a_1} }
  H_{2}^{0}[ a_1,x; m_1 , m_1 ; P] + \sum_{x\neq b_1} D_x {
    \spa{m_2}.{b_1} \over \spa{a_1}.{b_1} } H_{2}^{0}[ b_1,x; m_1, m_1
  ; P] \\+ D_{a_1} { \spa{m_2}.{a_1} \over \spa{b_1}.{a_1} }
  H^{0}_{2x}[ a_1 ; m_1 , m_1 ; P] + D_{b_1} { \spa{m_2}.{b_1} \over
    \spa{a_1}.{b_1} } H^{0}_{2x}[ b_1 ; m_1 ,m_1 ; P] \biggr),
\end{multline}
\begin{equation}
  \label{eq:bubcoeff2}
  \begin{split}
    c^{\NeqOne} [\{m_1,a_i\} ;\{m_2,b_i\} ] &= {1\over 2}{ \spa{m_1}.{m_2}^2
      \over (P^2)^{2} }\sum_{ P_L,P_R } C_{P_L} C_{P_R} \biggl(
    \\
    & \quad \quad \sum_{x\neq a_1} D_x { \spa{m_2}.{a_1} \over
      \spa{b_1}.{a_1} } H_{1,1}^{2}[ a_1;x; \{B_i\}^{\NeqOne} ; m_1 ;
    \{D_i\}^{\NeqOne} ; P] \ \\ &\quad + \sum_{x\neq b_1} D_x {
      \spa{m_2}.{b_1} \over \spa{a_1}.{b_1} } H_{1,1}^{2}[
    b_1;x;\{B_i\}^{\NeqOne} ; m_1 ; \{D_i\}^{\NeqOne} ; P]
    \\
    &\quad + D_{a_1} { \spa{m_2}.{a_1} \over \spa{b_1}.{a_1} }
    H^{2}_{2x}[ a_1 ;\{B_i\}^{\NeqOne} ; m_1 ;\{D_i\}^{\NeqOne} ; P]
    \\ & \quad + D_{b_1} { \spa{m_2}.{b_1} \over \spa{a_1}.{b_1} }
    H^{2}_{2x}[ b_1 ;\{B_i\}^{\NeqOne} ; m_1 ;\{D_i\}^{\NeqOne} ; P]
    \biggr),
  \end{split}
\end{equation}
where $P_L$ and $P_R$ are permutations of the positive helicity legs $\{a_i\}$ and $\{b_i\}$ respectively,
\begin{gather}
  \{B_i\}^{\NeqOne} =\{m_1,m_2,m_1\}, \quad
  \{D_i\}^{\NeqOne}  =\{ P|m_1\ra, P|m_2\ra \},\\
  C_{P_L} = {  \spb{n_L}.{m_1} \over \spa{n_L}.{m_1} \prod_{i=1}^{n_L-1} \spa{a_i}.{a_{i+1}} }, \quad 
   C_{P_R}= { \spb{n_R}.{m_2} \over \spa{n_R}.{m_2}\prod_{i=1}^{n_R-1} \spa{b_i}.{b_{i+1}} },
\end{gather}
and
$$
D_{x} = { \spa{m_2}.{x} \prod_{l=1}^{n_L-1} [ a_l|\tilde K_{l+1}|x\ra
  \prod_{k=1}^{n_R-1} [ b_k|\tilde K_{k+1}'|x\ra \over \prod_{y \neq
    x} \spa{x}.{y} } 
\equn
$$ 
where $\tilde K_p=k_{a_p}+\cdots k_{a_{n_L}}+k_{m_1}$ and  $\tilde K_p'=k_{b_p}+\cdots k_{b_{n_R}}+k_{m_2}$.
The
functions in \eqref{eq:bubcoeff1} and \eqref{eq:bubcoeff2}
of the form $H^N_{\{S\}}$ are the canonical forms~\cite{UsUnitarity}.
The index $N$ indicates the power of loop momenta present in the
cut. With increasing $N$ we find increasing complexity and increasing
powers of spurious denominators.
The simplest canonical form is
$$
H_1^0[A ; B ; P ] ={[A|P|B \ra \over [A|P|A\ra} \equn
$$
which is linear in the spurious singularity $[A|P|A\ra=2k_A\cdot P$.
It is convenient to define extensions,
$$
H_n^0[ A_i ; B_j ; P ]= \sum_i { \prod_{j=2}^n \spa{B_j}.{A_i} \over
  \prod_{j\neq i} \spa{A_j}.{A_i} } {\la B_1 | P | A_i ] \over \la A_i
  | P | A_i ] } \; , \;\;\; \spa{A_i}.{A_j} \neq 0 .  \equn
$$
and the special cases where $A_1=A_2=A$,
$$
{ H}^0_{2x}[ A, A ; B_1 ,B_2 ; P ]= { [A|P|B_1\ra [A|P|B_2\ra\over
  [A|P|A\ra^2 }.  \equn
$$
We also need the $H_{\{S\}}^1$,
\begin{equation}
  \shoveleft{
    H_0^1[B; D; P] = {1 \over 2} [D|P|B\ra, }
\end{equation}
\begin{equation}\begin{split}
    &H_1^1[ A; B_1,B_2; D ; P] = { P^2 \over 4 [A|P|A\ra^2 }\left(
      [D|A|B_1\ra[A|P|B_2\ra+ (B_1 \leftrightarrow B_2) \right) \\
    &\qquad +{ 1 \over 4 [A|P|A\ra } \left([D|P|B_1\ra [A|P|B_2\ra +
      (B_1 \leftrightarrow B_2) \right),
  \end{split}\end{equation}
\begin{equation}\begin{split}
    &{H}^1_{1,1}[A_1;A_2;B_1,B_2;B_3;D;P] = {
      \spa{A_1}.{B_3}\over\spa{A_1}.{A_2} } H_1^1[ A_1 ; B_1,B_2; D ;
    P] \\ & \qquad +{ \spa{A_2}.{B_3}\over\spa{A_2}.{A_1} } H_1^1[ A_2
    ; B_1,B_2; D ; P] +{ \spa{B_3}.{A_2} \spa{A_2}.{B_1} \over
      \spa{A_2}.{A_1} }\sum_{i\in P} \spb{D}.{i} H_{2x}^0[A_2 , i ,B_2
    ; P] \\ & \qquad +{ \spa{B_3}.{A_2} \spa{A_1}.{B_1} \over
      \spa{A_1}.{A_2} }\sum_{i\in P} \spb{D}.{i}
    H_{2}^0[A_1,A_2;B_2,i;P],
  \end{split}\end{equation}
\begin{equation}\begin{split}
    &H^1_{2x}[A;B_1,B_2;B_3;D;P] = { P^2 \over 3 [A|P|A\ra^3 } \left(
      [A|P|B_3\ra [A|P|B_1\ra [D|A|B_2\ra + (B_1 \leftrightarrow B_2)
    \right.  \\ & \qquad -2 \left. [A|P|B_1\ra [A|P|B_2\ra [D|A|B_3\ra
    \right) \\ & \qquad + { 1 \over 6 [A|P|A\ra^2 } \left( [A|P|B_1\ra
      [A|P|B_2\ra [D|P|B_3\ra +\Perm(\{B_i\}) \right),
  \end{split}\end{equation}
and the $H_{\{S\}}^2$
\begin{equation}\begin{split}
    & H_0^2[ B_1,B_2; D_1,D_2 ; P] ={1 \over 6}
    [D_1|P|B_1\ra[D_2|P|B_2\ra + (B_1 \leftrightarrow B_2),
  \end{split}\end{equation}
\begin{equation}\begin{split}
    &H_1^2[ A; B_1,B_2,B_3; D_1,D_2 ; P] = { (P^2)^2 \over 18
      [A|P|A\ra^3 } \left ( [D_1|A|B_1\ra [D_2|A|B_2\ra[A |P |B_3\ra+
      \Perm(\{B_i\}) \right) \\ &\qquad + { (P^2) \over 36 [A|P|A\ra^2
    } \left( [D_1|P|B_1\ra [D_2|A|B_2\ra[A |P |B_3\ra
      +\Perm(\{B_i\},\{D_i\})\right ) \\ &\qquad + { 1 \over 18
      [A|P|A\ra } \left( [D_1|P|B_1\ra[D_2|P|B_2\ra[A |P |B_3\ra+
      \Perm(\{B_i\}) \right),
  \end{split}\end{equation}
\begin{equation}\begin{split}
    & H_{1,1}^2[ A_1; A_2; B_1,B_2,B_3; C_4; D_1,D_2 ; P] = {
      \spa{C_4}.{A_1} \over \spa{A_2}.{A_1} }
    H_1^2[A_1,B_1,B_2,B_3,D_1,D_2,P] \\ &\qquad +{ \spa{C_4}.{A_2}
      \over \spa{A_1}.{A_2} } H_1^2 [A_2,B_1,B_2,B_3,D_1,D_2,P] \\
    &\qquad - { \spa{C_4}.{A_2}\spa{A_2}.{B_1}\over \spa{A_2}.{A_1} }
    H_{2x}^1[A_2,B_2,B_3,P|D_1], D_2,P] \\ &\qquad -{
      \spa{B_1}.{A_1}\spa{C_4}.{A_2}\over \spa{A_2}.{A_1} }
    H_{1,1}^1[A_1,A_2,B_2,B_3,P|D_1],D_2,P]\hfill,
  \end{split}\end{equation}
\begin{equation}\begin{split}
    & H_{2x}^2[ A; B_1,B_2,B_3; C_4; D_1,D_2 ; P] = \\ &\qquad {
      (P^2)^2 \over 9 [A|P|A\ra^4 } \left( [D_1|A|B_1\ra [D_2|A|B_2\ra
      [A|P|B_3\ra [A|P|C_4\ra +\Perm(\{B_i\}) \right) \\
    &\qquad - { (P^2)^2 \over 72 [A|P|A\ra^4 } \left( [D_1|A|C_4\ra
      [D_2|A|B_1\ra [A|P|B_2\ra [A|P|B_3\ra
      +\Perm(\{B_i\},\{D_i\}) \right) \\ &\qquad - { (P^2)
      \over 18 [A|P|A\ra^3 } \left( [D_1|A|C_4\ra [D_2|P|B_1\ra
      [A|P|B_2\ra [A|P|B_3\ra +\Perm(\{B_i\},\{D_i\}) \right)
    \\ &\qquad + { (P^2) \over 36 [A|P|A\ra^3 } \left( [D_1|P|C_4\ra
      [D_2|A|B_1\ra [A|P|B_2\ra [A|P|B_3\ra
      +\Perm(\{B_i\},\{D_i\}) \right) \\ &\qquad + { 1\over 18
      [A|P|A\ra^2 } \left( [D_1|P|B_1\ra [D_2|P|B_2\ra [A|P|B_3\ra
      [A|P|C_4\ra +\Perm(\{B_i\}) \right)\end{split}
\end{equation}

\section{Box Integral Functions} 
\label{IntegralFunctionsAp}
\def\L{\left(}
\def\R{\right)}
\def\tn#1#2{t^{[#1]}_{#2}}
\def\e{\eps}
\def\rg{r_\Gamma}

\def\Fn{n}
\def\Fs#1#2{F^{{#1}}_{}}
\def\Fone{\Fs{\rm 1m}}
\def\Feasy{\Fs{{\rm 2m}\,e}}
\def\Feasytwo{\Feasy}
\def\Fhard{\Fs{{\rm 2m}\,h}}
\def\Fhardtwo{\Fhard}
\def\Fthree{\Fs{\rm 3m}}
\def\Ffour{\Fs{\rm 4m}}
\def\cg{c_\Gamma}

\def\Lz{\mathop{\hbox{\rm L}}\nolimits_0}
\def\Li{\mathop{\hbox{\rm Li}}\nolimits}
\def\Ln{\mathop{\hbox{\rm Ln}}\nolimits}
\def\Li{\mathop{\hbox{\rm Li}}\nolimits}
\def\Kz{\mathop{\hbox{\rm K}}\nolimits_0}
\def\Mz{\mathop{\hbox{\rm M}}\nolimits_0}
\def\tr{\mathop{\hbox{\rm tr}}\nolimits}

\def\L{\left(}\def\R{\right)}
\def\LP{\left.}\def\RP{\right.}
\def\LB{\left[}\def\RB{\right]}
\def\LA{\left\langle}\def\RA{\right\rangle}
\def\LV{\left|}\def\RV{\right|}

The scalar box integral is,
$$
I_4 = -i \L4\pi\R^{2-\e} \,\int {d^{4-2\e}p\over \L2\pi\R^{4-2\e}}
\;{1\over p^2 \L p-K_1\R^2 \L p-K_1-K_2\R^2 \L p+K_4\R^2}\;,
\equn
$$
where $K_i$ is  the sum of the momenta of the legs attached to the $i$-th corner. 
If a single leg is attached then $K_i$ is null. The form of the integral depends upon the 
number of the $K_i$ which are non-null, $K_i^2\neq 0$. We often misname these {\it massive 
legs}. The integrals are functions of the non-zero $K_i^2$ and the invariants,
$$
S \equiv (K_1+K_2)^2 , \;\;\; T \equiv (K_2+K_3)^2 .
\equn
$$

The scalar box functions needed for our amplitudes,  expanded to
${\cal O}(\e^0)$, 
for the one-mass (with leg $4$ massive) and the two-mass-easy (with
legs $2$ and $4$ massive) are
\begin{align}
   I_{4}^{1{\rm m}} 
&=\ -{ 2 \rg  \over S T } \times  \Biggl( 
\ -{1\over\e^2} \Bigl[ (-S)^{-\e} +
(-T)^{-\e} - (-K_4^2)^{-\e} \Bigr] 
\notag\\
 &\ + \Li_2\left(1-{ K_4^2 \over S }\right)
  \ + \ \Li_2\left(1-{K_4^2 \over T}\right)
  \ +{1\over 2} \ln^2\left({ S  \over T}\right)\
+\ {\pi^2\over6}\  \Biggr) 
\notag\\
I_{4}^{2{\rm m}e}
& =\ -{ 2 \rg        \over ST  -K_2^2K_4^2  } \times  \Biggl( 
 - {1\over\e^2} \Bigl[ (-S)^{-\e} + (-T)^{-\e}
              - (-K_2^2 )^{-\e} - (-K_4^2 )^{-\e} \Bigr] 
\notag\\
&
\ +\ \Li_2\left(1-{ K_2^2 \over S }\right)
 \ +\ \Li_2\left(1-{ K_2^2 \over T}\right)
 \ +\ \Li_2\left(1-{ K_4^2 \over S }\right)
\notag \\
& +\ \Li_2\left(1-{ K_4^2 \over T }\right)
-\ \Li_2\left(1-{ K_2^2 K_4^2 
\over S T}\right)
   \ +\ {1\over 2} \ln^2\left({S/T}\right)\ 
\Biggr)
\end{align}
where
$$
\rg =  {\Gamma(1+\e)\Gamma^2(1-\e)\over\Gamma(1-2\e)},
\equn
$$
The {\it truncated} box-functions are these with the singularities
removed,

\begin{align}
   I_{4}^{1{\rm m, trunc\ }} 
=\ -{ 2 \rg  \over S T } \times  \Biggl( 
& \Li_2\left(1-{ K_4^2 \over S }\right)
  \ + \ \Li_2\left(1-{K_4^2 \over T}\right)
  \ +{1\over 2} \ln^2\left({ S  \over T}\right)\
+\ {\pi^2\over6}\  \Biggr) 
\notag\\
I_{4}^{2{\rm m}e, \rm trunc}
 =\ -{ 2 \rg        \over ST  -K_2^2K_4^2  } \times  \Biggl( 
& \Li_2\left(1-{ K_2^2 \over S }\right)
 \ +\ \Li_2\left(1-{ K_2^2 \over T}\right)
 \ +\ \Li_2\left(1-{ K_4^2 \over S }\right)
\notag \\
 +&\ \Li_2\left(1-{ K_4^2 \over T }\right)
-\ \Li_2\left(1-{ K_2^2 K_4^2 
\over S T}\right)
   \ +\ {1\over 2} \ln^2\left({S/T}\right)\ 
\Biggr)
\end{align}

The truncated zero-mass box (only necessary for the four-point amplitude) is
obtained by setting $K_4=0$ in the above expresion for  $I_{4}^{1{\rm m, trunc\ }}$.

\end{document}